\documentclass[hidelinks,breaklinks,12pt]{article}%
\usepackage[utf8]{inputenc}%
\usepackage{amsmath}%
\usepackage{amsfonts}%
\usepackage{amssymb}
\usepackage{graphicx}%
\usepackage[labelfont=bf]{caption} 
\usepackage{lmodern}%
\usepackage[backend=biber,
sorting=none]{biblatex}%
\renewcommand{\autocite}{ \cite}
\usepackage{makecell}

\title{Efficient Classical Simulation of the Deutsch-Jozsa
		and Simon's Algorithms}%
\author{Niklas Johansson \and Jan-Åke Larsson}

\newcommand{\ket}[1]{\ensuremath{\left\vert{#1}\right\rangle}}
\newcommand{\bra}[1]{\ensuremath{\left\langle{#1}\right\vert}}

\begin{document}

\maketitle
\begin{abstract}\noindent
	A long-standing aim of quantum information research is to understand
	what gives quantum computers their advantage. This requires separating
	problems that need genuinely quantum resources from those for which
	classical resources are enough. Two examples of quantum speed-up are the
	Deutsch-Jozsa and Simon's problem, both efficiently solvable on a quantum
	Turing machine, and both believed to lack efficient classical
	solutions. Here we present a framework that can simulate both quantum 
	algorithms efficiently,
	solving the Deutsch-Jozsa problem with probability 1 using
	only one oracle query, and Simon's problem using linearly many oracle
	queries, just as expected of an ideal quantum computer.  The presented 
	simulation framework is in turn efficiently simulatable in a classical 
	probabilistic Turing
	machine. This shows that the Deutsch-Jozsa and Simon's problem do not 
	require
	any genuinely quantum resources, and that the quantum algorithms show no
	speed-up when compared with their corresponding classical simulation.  
	Finally, this
	gives insight into what properties are needed in the two algorithms, and
	calls for further study of oracle separation between quantum and classical
	computation.
\end{abstract}\thispagestyle{empty}

\section*{Introduction}
\noindent
Quantum computational speed-up has motivated much research to build quantum
computers, to find new algorithms, to quantify the speed-up, and to separate
classical from quantum computation. One important goal is to understand the 
reason for quantum computational speed-up; to understand what resources are
needed to do quantum computation. Some candidates for such necessary resources
include superposition and interference\autocite{Feynman1982},
entanglement\autocite{Einstein1935}, nonlocality\autocite{Bell1964},
contextuality\autocite{Kochen1967,Larsson2012,Howard2014}, and the continuity of
state-space\autocite{Shor1996}. Here we will attempt to point out one such 
resource, sufficient for some of the known algorithms, and we will do this by 
simulation of the quantum algorithms in a classical Turing-machine. Simulation 
of small quantum-mechanical systems are routinely done on supercomputers 
world-wide, but classical Turing-machine simulations of this kind cannot be 
expected to be efficient. The complexity of a typical 
such simulation increases exponentially with the quantum system size. In other 
words, the complete quantum-mechanical behaviour of a quantum computation 
cannot be simulated efficiently\autocite{Feynman1982}. 

In this paper we consider the possibility that it might not be necessary to 
reproduce the complete quantum-mechanical behaviour to efficiently simulate
some quantum computation algorithms. In particular, we look at two so-called
\emph{oracle} problems: the Deutsch-Jozsa
problem\autocite{Deutsch,Deutsch1992} and Simon's
problem\autocite{Simon1994,Simon1997}, thought to show that quantum
computation is more powerful than classical computation.  There are
quantum algorithms that solve these problems efficiently, and here we present a 
framework that can simulate these quantum algorithms, 
\emph{Quantum Simulation Logic} (QSL), that itself can be efficiently simulated 
on a classical probabilistic Turing machine. This shows that no genuinely 
quantum resources are needed to solve these problems efficiently, but also 
tells us which properties are actually needed:
the possibility to choose between two aspects of the same system within which
to store, process, and retrieve information.

The quantum algorithm for the Deutsch-Jozsa problem has been used extensively
for illustrating experimental realizations of a quantum computer, while
Simon's algorithm served as inspiration for the later Shor's
algorithm\autocite{Shor1994}.  Both problems involve finding certain
properties of a function $f$, and this can be done efficiently in a quantum
computer given a quantum gate that implements the function, known as an
\emph{oracle}.  When using a classical-function oracle (an oracle that is 
only allowed to act on classical bits as input and giving classical bits as 
output), the
Deutsch-Jozsa problem has an efficient solution on a classical probabilistic
Turing machine\autocite{Deutsch,Deutsch1992}, but for Simon's problem it has
been proven\autocite{Simon1994,Simon1997} that no efficient solution exists. 
Our QSL simulation of Simon's algorithm may seem to contradict this theorem, 
since it runs on a classical probabilistic Turing machine, but there is no 
contradiction, because the function implemented by the QSL oracle is not a map from $n$ classical 
bits to $n$ classical bits --- which the theorem requires --- but a map from 
$n$ QSL systems to $n$ QSL systems. Thus, the theorem does not apply. This is exactly 
as in the quantum case, where the quantum oracle is a map from quantum systems 
to quantum systems, also avoiding the prerequisites of the theorem. Both the 
quantum and QSL oracles are richer procedures than the classical function 
oracle, and the main point of this paper is to clarify exactly how they are 
richer.

The paper is organized as follows: we first present the Deutsch-Jozsa problem 
and the quantum algorithm. We then construct a simulation of the quantum 
algorithm, in classical reversible logic. This is followed by a brief 
discussion of the black-box oracle paradigm, stressing that the simulation 
completely reproduces the behaviour of the Deutsch-Jozsa quantum algorithm in 
this paradigm. The final part of the paper is devoted to Simon's problem, 
quantum algorithm, and simulation, followed by the conclusions of the paper.

\section*{The Deutsch-Jozsa problem and quantum algorithm}

The Deutsch-Jozsa problem is the following: Suppose that you are given a
Boolean function $f:\{0,1\}^n\to\{0,1\}$ with the promise that it is
either constant or balanced. The function is constant if it gives the same
output (1 or 0) for all possible inputs, and it is balanced if it gives
the output 0 for half of the possible inputs, and 1 for the other
half. Your task is now to distinguish between these two
cases\autocite{Deutsch1992,Cleve}.  Given such a function, a classical Turing
machine can solve this problem by checking the output for $2^{n-1}+1$ values
of the input; if all are the same, the function is constant, and otherwise
balanced. A stochastic algorithm with $k$ randomized function queries gives a
bounded error probability\autocite{Deutsch1992} less than $2^{1-k}$, showing
that the problem is in the complexity class \textbf{BPP} (Bounded-error
Probabilistic Polynomial-time solvable problems).

\begin{figure}[t]
    \centering
    \includegraphics[scale=1]{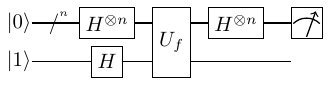}
    \caption{Quantum circuit for the Deutsch-Jozsa algorithm. This circuit uses 
    an $n$-qubit query-register prepared in the state $\ket{0}$, and an 
    answer-register prepared in $\ket{1}$. It proceeds to apply Hadamard 
    transformations to each qubit. The function $f$ is embedded in an 
    oracle $U_f$, and this is followed by another Hadamard transformation on 
    each query-register qubit. The measurement at the end will test 
    positive for $\ket{0}$ if $f$ was constant, and negative if $f$ was 
    balanced.}
    \label{fig:DJ}
\end{figure}

A quantum computer obtains the function as a unitary that implements the
function, a quantum oracle. The requirements on the oracle are simple: that it 
adds the function output onto the answer-register qubit for computational-basis 
inputs, 
and that the quantum phase is preserved in the process, so that
\begin{equation}
    \ket{x}\ket{y}\overset{U_f}{\mapsto}\ket{x}\ket{y\oplus f(x)}.
    \label{U_f}
\end{equation}
This corresponds to the classical single-input function query: if the 
answer-register was
flipped, then the function value is 1 for that input $x$. Given this quantum
oracle the Deutsch-Jozsa algorithm\autocite{Deutsch1992,Cleve} can solve the
problem with a single query (see Figure~\ref{fig:DJ}). In this algorithm
Hadamards are applied to the query- and answer-register states, creating a 
quantum superposition of all possible inputs to the oracle. When the oracle is
applied, this superposition is unchanged (modulo global phase) if the function is constant, and
changed to a different superposition if the function is balanced,
\begin{figure}[t]
    \centering
    \includegraphics[scale=1]{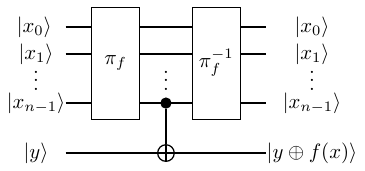}
    \caption{Oracle example for balanced functions.  The arbitrary
        permutation $\pi =\sum_x \ket{\pi(x)}\bra{x}$ of the computational basis
        states is constructed from CNOT and Toffoli gates \autocite{Nielsen}
        ($\pi^{-1}=\pi^{\dagger}$).  At the center is a CNOT gate from the most
        significant qubit $\ket{x_{n-1}}$ to the answer-register $\ket{y}$.  
        With $\pi$ as 
        the identity permutation, the oracle performs the balanced function 
        $f'$ that is 1 for all inputs with the most significant bit set.  Any 
        other balanced function $f(x)=f'(\pi_f(x))$ can now be generated by 
        choosing a suitable computational-basis permutation $\pi_f$.  For a 
        function that is constant zero, the CNOT gate is omitted, while the 
        constant one function should replace the CNOT with a Pauli-X gate 
        acting on the answer-register.}
    \label{fig:DJ-oracle}
\end{figure}
\begin{equation}
\sum_x\ket{x}\big(\ket0-\ket1\big)\overset{U_f}{\mapsto}
\sum_x\ket{x}\big(\ket{f(x)}-\ket{1\oplus 
f(x)}\big)=\sum_x(-1)^{f(x)}\ket x\big(\ket0-\ket1\big).
\label{U_f_-}
\end{equation}
This important phenomenon is sometimes called ``phase kick-back'' \cite{Cleve}.
A change can be detected by again applying Hadamards on the query-register and 
measuring in the computational basis. If the function is constant, the 
query-register state after the Hadamards is $\ket0$, and otherwise the 
query-register state is orthogonal to $\ket0$.  A computational-basis 
measurement will reveal this, in the ideal case with zero error probability, 
showing that the problem is in \textbf{EQP} (Exact or Error-free Quantum 
Polynomial-time solvable problems\autocite{Bernstein1997}). Thus, instead of an 
exponential number of calls to the function from a deterministic classical Turing machine, a 
single query of the quantum oracle is sufficient. 

\section*{Quantum Simulation Logic}
Having studied the Deutsch-Jozsa algorithm in some detail, we will now proceed 
to construct Quantum Simulation Logic (QSL), an approximate simulation of the 
quantum states and gates. 

To simulate qubits we will use pairs $(b_z,b_x)$
containing two classical bits, a ``computational'' bit $b_z$ and a ``phase'' bit
$b_x$.  These constitute the elementary systems of our QSL framework, the ``QSL
bits''. State preparation of a computational single qubit
state $\ket k$ is associated with preparation of $(k,X)$ where $X$ is a
random evenly distributed bit. Measurement in the computational basis is
associated with readout of the computational bit followed by randomization of
the phase bit, somewhat like the uncertainty relation or really measurement
disturbance as seen in quantum mechanics. Accordingly, measurement of the phase
bit will be followed by a randomization of the computational bit. These
constructions of state preparation and measurement prohibits exact
preparation and readout of the system; the upper limit is one bit of
information per QSL bit $(b_z,b_x)$. 

\begin{figure}[t!]
	\center
	\begin{tabular}{l l}
		\makecell{
		\raisebox{.6cm}{\textbf{A.}}\includegraphics[scale=1]{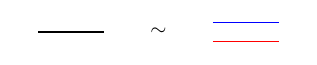}\\
		\raisebox{.6cm}{\textbf{B.}}\includegraphics[scale=1]{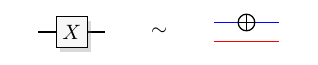}\\
		\raisebox{.6cm}{\textbf{C.}}\includegraphics[scale=1]{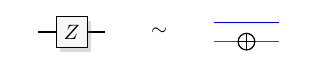}\\
		\raisebox{.6cm}{\textbf{D.}}\includegraphics[scale=1]{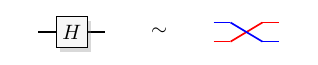}\\
		}
		&
		\makecell{
		\raisebox{1.9cm}{\textbf{E.}}\includegraphics[scale=1]{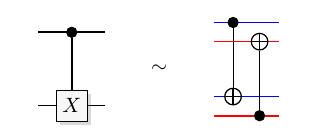}\vspace{.5cm}\\
		\raisebox{1.6cm}{\textbf{F.}}\includegraphics[scale=1]{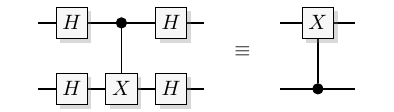} 
				}

	\end{tabular}
	\caption{QSL circuitry. \textbf{A.}~Identity operation, a qubit is simulated by
		a pair of classical bits (computational bit in blue, phase bit in red).
		\textbf{B.}~QSL construction of a NOT gate. \textbf{C.}~QSL construction of a
		Z-gate. \textbf{D.}~QSL construction of a Hadamard gate. \textbf{E.}~QSL construction of a CNOT gate.
		\textbf{F.}	Identity valid both for qubit and QSL CNOT gates. }
	\label{fig:Aqubits}
\end{figure}

It is relatively simple to simulate the single qubit gates used in the quantum
algorithms: an $X$ gate inverts the computational bit
(preserving the value of the phase bit $b_x$), a $Z$ gate inverts the phase bit
(preserving $b_z$), and a $H$ gate switches the computational and phase
bits, see Figure~\ref{fig:Aqubits}B--D. These are constructed so that the resulting gates obey the quantum
identities $XX=ZZ=HH=I$ and $HZH=X$. It is also possible to define QSL
$Y$ and ``phase'' gates, but we will not discuss these and their associated
identities here as they are not needed for the present task. A simple circuit
is to prepare the state $\ket0$, apply a $H$ gate, and measure in the
computational basis. This will give a random evenly distributed bit as output.

A system of several qubits can now be associated with a system of several
QSL bits containing two internal bits each. A QSL simulation of the quantum Controlled-NOT (CNOT)
gate can be constructed from two classical reversible-logic CNOTs. One
classical CNOT connects the computational bits in the ``forward''
direction, while the other connects the phase bits in the ``reverse''
direction, directly implementing the phase 
kick-back, see Figure~\ref{fig:Aqubits}E. This again is constructed to enable 
use of the same identities as apply for the quantum CNOT, see 
Figure~\ref{fig:Aqubits}F.

Using QSL to simulate a generic qubit circuit is done as follows:
\begin{enumerate}
	\item \textbf{State preparation.} Prepare $n$ QSL systems (pairs of 
	classical bits) in states $\ket{0}$ or $\ket{1}$ (using a random value for 
	the phase bit).
	\item \textbf{Transformation.} Apply QSL gates (as described above) in an 
	array of the same form as the quantum circuitry.
	\item \textbf{Measurement.} Measure in the computational basis (readout the 
	computational-bit value and randomize the phase bit).
\end{enumerate}

The QSL framework is 
inspired by a construction of Spekkens\autocite{Spekkens} that captures many, 
but not all, properties of quantum mechanics.  So far, the presented framework 
is completely equivalent to Spekkens' model (the $H$ gate appears in a paper by
Pusey\autocite{Matthew}).  Both frameworks are efficiently simulatable on a
classical Turing machine (see above and Ref.~\cite{Spekkens}). Furthermore,
both have a close relation to stabilizer states and Clifford group
operations\autocite{Matthew}, and perhaps more interestingly, both frameworks
capture some properties of entanglement, enabling protocols like super-dense
coding and quantum-like teleportation, but cannot give all consequences of
entanglement, most importantly, cannot give a Bell inequality violation.

\section*{Simulation of the Deutsch-Jozsa algorithm in Quantum Simulation Logic}

The QSL versions of the quantum algorithm will use the
same setup as in Figure \ref{fig:DJ}. For $n=1$ the oracle for the balanced
function $f(x)=x$ is a CNOT, which means that Figure~\ref{fig:Aqubits}E
contains the whole Deutsch-Jozsa algorithm when using the prescribed initial
state $\ket0\!\ket1$.  With this initial state, the query-register measurement
result would be 1, so that this oracle gives the same output from the
QSL algorithm as from the quantum algorithm. Indeed, all oracles
for $n=1$ and $n=2$ only use CNOT and $X$ gates, so their simulation will give the same behaviour as the quantum oracles. For $n=1$ and $n=2$, the Deutsch-Jozsa algorithm is
already known to have an implementation that does not rely on quantum
resources\autocite{Collins}, and also, the gates used so far can be found
within the stabilizer formalism, and here the Gottesmann-Knill theorem tells us 
that they can be simulated (efficiently)\autocite{Gottesman1998}.

\begin{figure}[t!]
	\center
	\begin{tabular}{l l}
		\makecell{
			\raisebox{2.8cm}{\textbf{A.}}\includegraphics[scale=1]{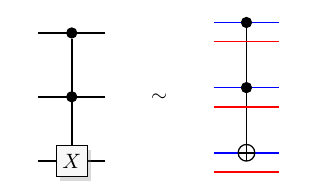}\\
		}
		&
		\makecell{
			\raisebox{2.4cm}{\textbf{B.}}\includegraphics[scale=1]{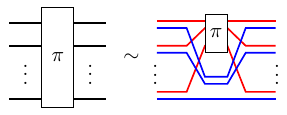}\\
		}
	\end{tabular}
	\caption{QSL circuitry. \textbf{A.}~QSL construction of a Toffoli gate. 
		\textbf{B.}~QSL	construction of a general permutation.}
	\label{fig:QSL_Toffoli}
\end{figure}

However, for $n\ge3$, the Deutsch-Jozsa oracle needs the Toffoli gate, and
since the Toffoli gate cannot be efficiently simulated using the stabilizer
formalism\autocite{Gottesman1998}, nor is present in Spekkens'
framework\autocite{Spekkens,Niklas}, it has so far been believed that the
Deutsch-Jozsa algorithm does not have an efficient classical simulation.
Here, we need to point out that our task is not to create exact Toffoli gate
equivalents, or even simulate the full quantum-mechanical system as such. It
suffices to give a working efficient QSL version of the
Deutsch-Jozsa algorithm. We therefore choose not to represent Toffoli gates
exactly, but design the gate so that it implements a classical Toffoli over the
computational bits, and some other gate in the phase bits. For simplicity we
choose the identity map for the phase bits, see Figure~\ref{fig:QSL_Toffoli}A.

\begin{figure}[t]
	\begin{center}
		\includegraphics[scale=1]{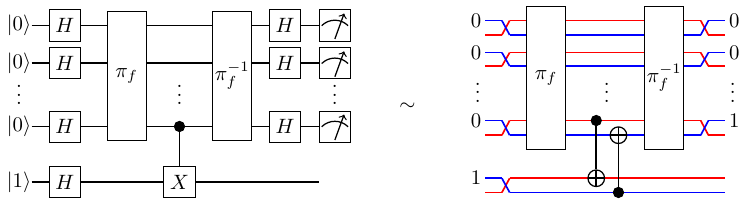}
	\end{center}
	\caption{Translation of the Deutsch-Jozsa quantum algorithm for a balanced 
	function into the QSL framework. The phase kick-back of the CNOT changes 
	the most significant phase bit of the input register, while the $\pi$ gates 
	does not change the phase bits at all. The final Hadamard gate will make 
	the computational base measurement reveal the changed value. }
	\label{fig:DJ-translation}
\end{figure}	

Building a general computational-state permutation $\pi$ from QSL Toffoli gates 
will result in a mapping that only changes the computational bits, leaving the 
phase-bits unchanged, see Figure~\ref{fig:QSL_Toffoli}B. The quantum oracle of 
Figure~\ref{fig:DJ-oracle} can immediately be translated into an oracle within 
the QSL framework, see Figure~\ref{fig:DJ-translation}. The balanced-function 
oracle leaves the phase bits unchanged except for the one belonging to the most 
significant QSL bit in the query-register. A constant-function oracle leaves 
all phase bits unchanged. Since Hadamard gates are included before and after 
the oracle, measurement of the computational bits will reveal the oracle's 
effect on the phase bits of the query-register,  solving the problem by only 
one oracle query.

This QSL algorithm uses equally many QSL bits and QSL gates as the
quantum algorithm uses qubits and quantum gates. The time and space complexity
are therefore identical to the complexity of the quantum algorithm. The
QSL algorithm can be efficiently simulated on a classical
probabilistic Turing machine, using two classical bits for each simulated qubit 
and at most twice as many classical reversible gates as quantum gates. The
error probability is 0, and in fact, the same correct output is generated for
all possible random values in the probabilistic machine, so that the machine's
random value could be replaced with a constant value without destroying the
simulation. Therefore, this QSL algorithm can be simulated on a
classical \emph{deterministic} Turing machine, showing that, relative to the
oracle, the Deutsch-Jozsa problem is in \textbf{P} (Polynomial time solvable
problems).

The problem is solvable in one QSL oracle query because of the oracle's effect
on the phase bits of the query-register, reproducing the 
quantum-mechanical phase kick-back. The phase bits of the
query-register reveals the inner structure of the
oracle. One may question if it is fair to compare our construction to a
\textit{classical function}, since it is clear that this construction results 
in an oracle that allows one to retrieve either function output or function
structure. But then one must also question the comparison of the quantum
algorithm to the classical-function algorithm, since the quantum oracle also
allows one to retrieve either function output or function structure.
On the other hand, the comparison between the QSL simulation and the 
\textit{quantum
algorithm} really \textit{is} a fair comparison since the choice of what to 
reveal is
present in both oracles; both oracles actually result in the same algorithm.
And in this comparison, there is no quantum advantage with respect to the
classical simulation of QSL.

\section*{Equal requirements on quantum and QSL oracles}

It has been suggested that the construction of the QSL oracle invalidates the 
simulation: the above-used simple construction of the QSL oracle contains a 
single CNOT between query- and answer-registers for a balanced function, and 
does not 
for a constant function. It seems simple to check for presence/absence of the 
CNOT in the gate array. With the present design of QSL, CNOT presence or 
absence is in fact required for the algorithm to work. There are two points to 
make here.

First, we must stress that the Deutsch-Jozsa problem is stated in the 
\textit{black-box oracle paradigm}\autocite{Deutsch1992,Cleve}. The quantum 
speedup itself is only obtained in the black-box oracle paradigm: if the 
explicit gate construction of the function is available (in a ``white-'' or 
``transparent-box paradigm''), the same simple solution may be usable in the 
fully quantum case, for example in the oracle construction of Figure 
\ref{fig:DJ-oracle}. Oracles built from this design, using classical reversible 
logic, quantum, or QSL gates will \textit{all} indicate that the function is 
balanced or constant from  presence or absence of the CNOT, without the need 
for any function query. For this choice of (transparent-box) ``oracle'' and 
many others including most or even all quantum-experimental implementations to 
date, there will be no quantum speedup when comparing to this simple 
``solution'' of the problem. However, the appropriate comparison between 
quantum and classical algorithms is that between algorithms that use only the 
inputs and outputs of the function or oracle, not the structure of the gate 
array at hand\autocite{Deutsch1992,Cleve}, corresponding to the black-box 
oracle paradigm. Only in this paradigm does the generic quantum speedup remain. 
And in this paradigm, the QSL simulation reproduces the quantum behaviour.

Second, the \textit{essential requirement} on these black-box oracles is 
not structural, neither in quantum computation nor in QSL simulation. The 
essential requirement is presence of phase kick-back. For a quantum oracle, the 
standard way to enable phase kick-back is to enforce phase preservation as 
hinted in eqn.~(\ref{U_f}), or more explicitly
\begin{equation}
\sum_{x,y}c_{x,y}\ket x \ket y \overset{U_f}{\mapsto}\sum_{x,y}c_{x,y}\ket x 
\ket{y\oplus f(x)}.\label{U_f_phase}
\end{equation}
It should be stressed that unitarity is not enough of a requirement to preserve 
phase in the sense of eqns.~(\ref{U_f})--(\ref{U_f_phase}), even if the 
functional map from query- to answer-register is enforced. There are many 
unitary maps that obey the functional relation in the computational basis, but 
do not preserve the phase relation. One simple but important example is the 
unitary $U_f'$ defined by
\begin{equation}
\sum_{x,y}c_{x,y}\ket x \ket y\overset{U_f'}{\mapsto}\sum_{x,y}(-1)^{f(x)} 
c_{x,y}\ket x \ket{y\oplus f(x)}.
\label{U_f_prime}
\end{equation}
This can easily be constructed from $U_f$ by adding $Z$ gates on both sides 
on the answer-register system, corresponding to adding identity gates in 
classical 
reversible logic. The  $U_f'$ unitary gives the correct map from function input 
to output in the computational basis, but if used as the quantum function 
oracle, the Deutsch-Jozsa algorithm input would give
\begin{equation}
\sum_x\ket x\big(\ket0-\ket1\big)\overset{U_f}{\mapsto}
\sum_x(-1)^{f(x)}\ket x\big(\ket{f(x)}-\ket{1\oplus 
    f(x)}\big)=\sum_x\ket x\big(\ket0-\ket1\big),
\label{U_f_-p}
\end{equation}
resulting in no change to the state used in the algorithm. Here, the output of 
the algorithm is 0 independent of the properties of the function. In other 
words, there is no phase kick-back. With this oracle, the quantum algorithm 
cannot distinguish between constant and balanced functions, and one will need 
to resort to the classical algorithms. It may be tempting to try to devise 
modified quantum algorithms that adjust to the type of oracle received, but 
alas, there are many unitary maps that obey the functional relation but not the 
phase relation required in the Deutsch-Jozsa algorithm. Testing for the type of 
oracle is nontrivial and even with a discrete set of phases, the set of 
possible unitaries is exponential in size. The phase relation that enables the 
quantum speedup restricts this set to a minimum; the remaining oracle 
unitaries all enable phase kick-back. 

The QSL framework presented above is deliberately chosen as simple as possible, 
to make it clear what properties of quantum mechanics is really needed for the 
two algorithms under study (Simon's algorithm follows below). A direct 
consequence of this simplicity is that the phase relation requirement of 
quantum computation, that enables phase kick-back, is translated into a simple 
requirement on the QSL oracles, presence or absence of a CNOT between query- 
and answer-register. In quantum computation, the restriction that enables phase 
kick-back selects a small subset of all oracle constructions that give the 
correct functional map for the computational basis, and in QSL, exactly in the 
same way, the restriction that enables phase kick-back selects a small subset 
of all QSL constructions that give the correct functional map for the 
computational bits. 

The requirements are exactly the same on quantum and QSL oracles alike: a 
black-box oracle that enables phase kick-back. The above example proves 
existence of black-box oracles that enable phase kick-back for the full range 
of possible functions, both in the quantum and QSL frameworks. And in the 
black-box oracle paradigm, the central issue is \textit{existence} of oracles 
with the appropriate properties, not their internal structure, the latter not 
being available to the algorithms in question. 

\section*{Simon's problem}

Another more complicated problem that can be solved faster with quantum 
computation is Simon's problem, to decide whether a function
$f:\{0,1\}^n\to\{0,1\}^n$ is one-to-one or two-to-one invariant under the
bitwise exclusive-OR operation with a secret string $s$ (or equivalent,
invariant under a nontrivial XOR-mask $s$~\cite{Simon1994,Simon1997}). In both
cases, $f(x)=f(x')$ if and only if
$x'=x\oplus{s}$ (bit-wise addition modulo 2); if $s=0$ then $f$ is one-to-one
and otherwise $f$ is two-to-one.  When using the classical function as an 
oracle, a classical probabilistic Turing machine needs an exponential number of 
oracle evaluations for this task due to a theorem by Simon 
\autocite{Simon1994,Simon1997}, showing that relative
to this oracle, Simon's problem is not in~\textbf{BPP}.

\begin{figure}[t!]
    \hspace*{1cm}\raisebox{0.9cm}{\textbf{A.}}\hspace{0cm}\includegraphics[scale=1]{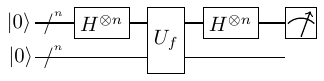}\\[1.2em]
    \hspace*{1cm}\raisebox{1.5cm}{\textbf{B.}}\includegraphics[scale=1]{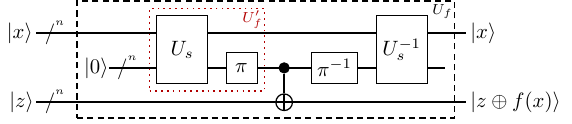}\\[1.2em]
    \hspace*{1cm}\raisebox{1cm}{\textbf{C.}}\includegraphics[scale=1]{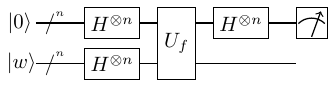}
    \\
    \caption{Circuits for the quantum subroutine of Simon's
        algorithm. \textbf{A.}~The subroutine itself. The $n$-qubit query- and 
        answer-registers are prepared in the state $\ket{0}\!\ket{0}$. Apply Hadamard
        transformations to the query-register, the function $f$ embedded in an
        oracle $U_f$, and finally another Hadamard. The measurement at the end 
        will
        give a random bit sequence $y$ such that $y\cdot{s}=0$ (mod 2). 
        %
        \textbf{B.}~Example of oracle construction. The oracle $U_f$ uses an 
        internal
        $n$-qubit ancilla, and the CNOT denotes a string of bit-wise CNOTs from 
        each
        ancilla bit to the corresponding answer-register bit. The unitary $U_s$ 
        is
        constructed from CNOTs\autocite{Tame2014} (see the main text) so that it
        implements one representative function for each specific secret string 
        $s$, and the
        permutation $\pi$ gives access to all functions $f$. The dotted gate
        combination $U_f'$ implements $\ket x\!\ket0\mapsto\ket x\!\ket{f(x)}$ 
        as
        required in Simon\autocite{Simon1994,Simon1997}, and the additional
        circuitry in $U_f$ is there to map $\ket{x}\!\ket{z}$ to
        $\ket x\!\ket{z\oplus f(x)}$ and reset the 
        ancilla\autocite{Bennett1973}.
        \textbf{C.}~The modified subroutine for the deterministic algorithm. The
        $n$-qubit query- and answer-register are prepared in the state
        $\ket{0}\!\ket{w}$. The circuit applies Hadamard transformations to the 
        query-
        \emph{and answer-registers}, the function $f$ embedded in an oracle 
        $U_f$,
        and finally another Hadamard before measurement.}
    \label{fig:Simon}
 \end{figure}

\begin{figure}[h!]
	\begin{center}
		\includegraphics[scale=1]{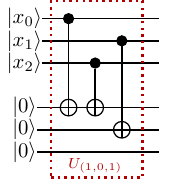}
	\end{center}
	\caption{Example construction of $U_s$ when $s=(1,0,1)$, using the 
	basis $\{v^k\}=\{(1,0,1),(0,1,0)\}$ which spans	the binary vector space 
	orthogonal to~$s$ (note that the scalar product is defined mod 2). The 
	construction uses three CNOTs, one from the first query-register qubit to 
	the first ancilla qubit, one from the third query-register qubit to the 
	first ancilla qubit, and one from the second query-register qubit to the 
	second ancilla qubit,}
	\label{fig:example_simons}
\end{figure}

Simon's quantum algorithm can distinguish these two cases in an expected
linear number of oracle queries\autocite{Simon1994,Simon1997}, showing that
the problem is in \textbf{BQP} (Bounded-error Quantum Polynomial
time\autocite{Bernstein1997}). A circuit diagram of the quantum subroutine and
one realization of the oracle used in Simon's algorithm is shown in Figure
\ref{fig:Simon}A--B.  An explicit construction of the gate $U_s$ can be
obtained by choosing a basis $\{v^k\}$ for the part of the binary vector space
orthogonal to~$s$. If $s=0$, the basis consists of $n$ vectors, otherwise
$n-1$ vectors. For every entry $v_j^k=1$, connect a CNOT from query-register
bit $j$ to ancilla-register $k$ (in realizations, it is beneficial to choose
${v^k}$ to minimize the number of CNOTs\autocite{Tame2014}). The permutation
after $U_s$ enables all possible functions.  See 
Figure~\ref{fig:example_simons} for a simple example.

The quantum subroutine output is a random bit vector $y$ such that
$y\cdot{s}=0$ mod 2 (bit-wise dot product), uniformly distributed over
all such bit vectors $y$.  After an expected number of iterations that is
linear in $n$, the subroutine will have produced $n-1$ linearly
independent values of $y$\autocite{Simon1994,Simon1997}, so that the
resulting linear system of equations for $s$ can be solved, giving one
non-trivial solution $s^*$. If the function is one-to-one, this
solution is just a random bit sequence, but if the function is
two-to-one the solution is the actual secret string $s$. Querying the
function for $f(0)$ and $f(s^*)$ will give equal values if the
function is two-to-one, and unequal values if the function is
one-to-one, solving Simon's problem. Note that as a consequence of
solving Simon's problem one also obtains the secret string~$s$.

The QSL versions of the subroutine and oracle is again obtained by
substituting the quantum gates with their corresponding QSL gates. Inserting the
QSL CNOT into $U_s$ gives the explicit map
$(x,p),(0,a)\overset{U_s}{\mapsto}\big(x,p\oplus g'(a)\big),\big(f'(x),a\big)$ 
where
\begin{equation}
  \label{eq:1}
  \begin{split}
    f_j'(x)&=x\cdot v^j\text{ (mod }2)=f_j'(x\oplus s),\\
    g'_j(a)&=\sum_k a_k v_j^k\text{ (mod }2).
  \end{split}
\end{equation}
The permutations $\pi$ and $\pi^{-1}$ do not influence the phase, so
the complete oracle $U_f$ is the map
$(x,p),(z,w)\mapsto\big(x,p\oplus g(w)\big),\big(z\oplus f(x),w\big)$
where
\begin{equation}
  \label{eq:2}
  \begin{split}
    f(x)&=\pi(f'(x))=f(x\oplus s),\\
    g(w)&=g'(a)\oplus g'(w\oplus a)=g'(w)
  \end{split}
\end{equation}
By the use of Hadamard gates, the simulation of Simon's subroutine sets the
query-register phase entering the oracle to $p=0^n$, while the
answer-register phase $w$ will be uniformly distributed. Measurement of the
computational bits after the final Hadamard will therefore give a random bit
vector $y=g(w)$ that is orthogonal to~$s$, uniformly
distributed over the possible values. This reproduces the quantum predictions
exactly, showing that the simulation will work with this subroutine to solve
Simon's problem with the same expected linear number of iterations in $n$. 
Again, efficient simulation of the QSL framework on a classical
probabilistic Turing machine shows that, relative to the oracle, Simon's
problem is in \textbf{BPP}.

This seems to contradict the above-mentioned theorem of 
Simon\autocite{Simon1994,Simon1997}, that states that relative to a 
classical-function oracle, Simon's problem is not in~\textbf{BPP}. However, 
there is no contradiction, because the QSL oracle is not a map from $n$ classical 
bits to $n$ classical bits, which the theorem requires, but a map from $n$
QSL systems to $n$ QSL systems. Thus, the theorem does not apply, and there is no 
contradiction: relative to the new oracle, Simon's problem is not stopped from 
being in \textbf{BPP}. This is because of the additional resource available in 
QSL systems, the choice between two aspects of the same system, the 
``computational'' or the ``phase'' bit, within which to store, process, and 
retrieve information.  It is of course crucial that the QSL framework itself 
can be efficiently simulated on a classical probabilistic Turing machine.

The derandomized quantum algorithm for Simon's problem\autocite{Brassard1997,
	Mihara2003} (with zero error probability), that shows that Simon's problem
relative to the quantum oracle is in \textbf{EQP}, is not immediately usable
because it incorporates a modified Grover search\autocite{Grover1996} which is
not known to have an implementation within the QSL framework. The quantum
algorithm is intended to remove already seen $y$ and, crucially, prevent
measuring $y=0$ which would be
a failed iteration of the algorithm. The modified Grover search is
presented\autocite{Brassard1997} as applied to the output of $U_f$, but since
the final step of the Grover search is again $U_f$, the whole combination can
be viewed as $U_f$ preceded by another quantum algorithm. This other quantum
algorithm, in a sense, finds an input to $U_f$ such that the coefficient of
the term $\ket0$ of the output is~0.

Since we do not have a Grover equivalent in our framework, we would like a
simpler solution. One way to achieve this is to prepare the initial
answer-register state as the Hadamard transformation of a chosen state
$\ket{w}$, see Figure \ref{fig:Simon}C. Iterating $\ket{w}$ over a basis for
the bit-vector space will produce $n$ values of $y$ that span the bit vector
space orthogonal to~$s$. An example is to use the standard basis, for which the 
outputs can be calculated through
eqns.~(\ref{eq:1}--\ref{eq:2}) to be the individual $v_i$, in order. Either 
this spans the whole space ($s=0$), or gives a
linear system of equations that has one non-trivial solution, equal to
$s$. This QSL algorithm gives deterministic correct outputs just as
the QSL Deutsch-Jozsa algorithm, and can therefore also be efficiently
simulated on a classical deterministic Turing machine, showing that relative
to the oracle, Simon's problem actually lies within~\textbf{P}.

This linear search procedure works in our setting because the QSL permutation
gate $\pi$ does not influence the phase bits.  In the
quantum-mechanical case, the quantum gate implementing $\pi$ does change the 
phase information, e.g., in the simple case when
the quantum $\pi$ is a CNOT. A more convoluted example that does not
correspond to a simple modification of the phase information is when the
quantum $\pi$ is a (non-stabilizer) Toffoli gate.  Thus, the derandomized
quantum algorithm\autocite{Brassard1997} requires the more advanced modified
Grover's search algorithm.  However, note that if the oracle is the smaller
gate combination $U_f'$ in Figure~\ref{fig:Simon}B that obeys only the
original\autocite{Simon1994,Simon1997} requirement
$\ket{x}\!\ket{0}\mapsto\ket{x}\!\ket{f(x)}$ (for $z=0$), then the quantum
version of the simpler algorithm proposed above will work as well, avoiding
the modified Grover search.

Also here the simulation makes it clear that Simon's quantum oracle is
a richer procedure than the classical function oracle, since it
allows for the retrieval of more information than just the function output.
Choosing to view the oracle operation from the computational basis reveals the
function output, while choosing the phase basis reveals function structure, in
the form of output bit vectors orthogonal to~$s$.
Also here, the comparison between the QSL simulation and the quantum
algorithm \textit{is} a fair comparison since the choice of what to reveal is
present in both oracles; both oracles actually result in the same algorithm.
And also in this comparison, there is no quantum advantage with respect to the
classical simulation of QSL.

\section*{Conclusions}

In conclusion, we have devised QSL equivalents of the Deutsch-Jozsa and Simon's 
quantum algorithms. In the quantum algorithm, you are given a black-box quantum 
oracle, a unitary gate that implements $f$ by acting on qubits, your task is to 
distinguish two or more families of functions.  In the presented QSL 
algorithms, you are given a black-box QSL gate that implements $f$ by acting on 
QSL systems, and the same task.  Using the QSL framework, we obtain the same 
success probability and the same time and space complexity as for the quantum 
algorithms, and these QSL algorithms are in turn efficiently simulatable in 
classical Turing machines. (Implementations that can be run with very large 
inputs can be constructed, given that the permutations are implemented as random
permutations.) Therefore, the Deutsch-Jozsa and Simon's algorithms cannot any 
more be used as evidence for a relativized oracle separation between 
\textbf{EQP} and \textbf{BPP}, or even \textbf{P}. Both problems are, relative 
to the oracles, in fact in \textbf{P}.

Apparently, the resource that gives these quantum algorithms their power is
also available in QSL, but not when using an classical-function oracle.  In QSL,
each qubit is simulated 
by two classical bits in a classical probabilistic Turing machine, and
in the above oracles one of these bits is used for computing the function,
while the other is changed systematically when $U_f$ is
applied. This change of phase information can then be revealed by
choosing to insert Hadamards before and after the oracle.  This choice is
present in both quantum systems and our simulation, and enables
revealing either function output or function structure, as desired.  Our
conclusion is that the needed resource is the possibility to choose between
two aspects of the same system within which to store, process, and retrieve
information.

While we still believe that quantum computers \emph{are} more powerful than
classical computers, the question arises whether oracle separation really can
distinguish quantum computation from classical computation. There are many other
examples of quantum-classical oracle
separation,\autocite{Fan2007,Alagic2007,NISTOracular} and some simple cases
can be conjectured to have efficient simulations in the QSL framework (e.g.,
$3$-level systems\autocite{Spekkens} for the three-valued Deutsch-Jozsa
problem\autocite{Fan2007}), but in general the question needs further study.
It \emph{is} possible that the technique can be used for direct
computation using other quantum algorithms\autocite{Grover1996,Shor1994}, but
this will take us out of the oracle paradigm.  Also, general quantum
computation has been conjectured to use genuinely quantum properties such as
the continuum of quantum states, or
contextuality\autocite{Larsson2012,Howard2014}, which both are missing from
the QSL framework\autocite{Spekkens}.  In any case, neither the
Deutsch-Jozsa nor Simon's algorithm needs genuinely quantum mechanical
resources.

\section*{Acknowledgments}

The authors would like to thank Peter Jonsson, Chris Fuchs, Markus Grassl, and
David Mermin for discussions on the subject.  The project has been supported by
the Foundational
Questions Institute (\href{www.fqxi.org}{FQXi}) through the Silicon Valley
Community Foundation.

\section*{References}
\printbibliography[heading=subbibliography,
  title={\null}]

%
%
%
%
%
%
%
%
%

\end{document}